\documentclass[aps,prb,twocolumn,floatfix,floats,showpacs,superscriptaddress,raggedbottom]{revtex4}

\usepackage{graphicx,latexsym}
\usepackage{dcolumn}
\usepackage{amsmath,amssymb,bm}% ams math, ams symbol, bold math
\usepackage{epsfig}
\usepackage[english]{babel}

\def\fig#1{Fig.\ \ref{#1}}
\def\sec#1{Sec.\ \ref{#1}}
\def\bib{\bibitem}

\begin{document}

\title{Time-dependent magnetotransport of a wave packet\\
in a quantum wire with embedded quantum dots}

\author{Gunnar Thorgilsson}
\affiliation{Science Institute, University of Iceland, Dunhaga 3,
IS-107 Reykjavik, Iceland}
\author{Chi-Shung Tang}
\email{cstang@sinica.edu.tw}%
\affiliation{Research Center for Applied Sciences, Academia Sinica,
Taipei 11529, Taiwan}
\author{Vidar Gudmundsson}
\email{vidar@raunvis.hi.is}%
\affiliation{Science Institute, University of Iceland, Dunhaga 3,
IS-107 Reykjavik, Iceland}

%----------------------------------------------------------------

\begin{abstract}
We consider wave packet propagation in a quantum wire with either an
embedded antidot or an embedded parallel double open quantum dot
under the influence of a uniform magnetic field.  The
magnetoconductance and the time evolution of an electron wave packet
are calculated based on the Lippmann-Schwinger formalism. This
approach allows us to look at arbitrary embedded potential profiles
and illustrate the results by performing computational simulations
for the conductance and the time evolution of the electron wave
packet through the quantum wire.  In the double-dot system we
observe a long-lived resonance state that enhances the spatial
spreading of the wave packet, and quantum skipping-like trajectories
are induced when the envelop function of the wave packet covers
several subbands in appropriate magnetic fields.
\end{abstract}

\pacs{72.10.-d, 73.21.Hb, 73.23.-b, 75.47.-m}
% 73.23.-b     Electronic transport in mesoscopic systems
% 72.10.-d    Theory of electronic transport; scattering mechanisms
% 73.21.Hb    Quantum wires
% 73.23.Ad    Ballistic transport
% 75.47.-m    Magnetotransport phenomena; materials for magnetotransport

\maketitle

%------------------------------------------------------------------------
\section{INTRODUCTION}

Recent progress in nanotechnology enables us to fabricate various
types of quantum systems embedded in nanostructures in which the
charge carriers behave coherently.\cite{nanostructure}   Electronic
transport in these mesoscopic or nanoscopic size systems is
phase-coherent, and the universal quantization of the dc conductance
is one of the well-known features that was measured in various
semiconductor structures.\cite{Wees88-Koester93-Scappucci06} The
low-temperature behavior of the conducting electrons becomes
dominated by quantum interference effects.  For example, a single
impurity allows the electrons to make coherent elastic intersubband
transitions forming quasibound states nearby the threshold of a
subband bottom.\cite{Chu89-Bagwell90-Bardarson04} One of the
advantages of electronic transport is its tunability by applying
external magnetic
fields.\cite{Berggren89-Tang-Vidar,Olendski05,Rogge06,Koonen00,%
Lofgren06,Aidala07,Gabelli06-07}
Transport properties are affected by the nature of current-carrying
states in the leads connecting these structures to electron
reservoirs.   The electronic transport under influence of an
external magnetic field has been utilized in several aspects such as
probing impurities in nanostructures under depleted
conditions,\cite{Koonen00} studying magnetoconductance
fluctuations,\cite{Lofgren06} imaging magnetic focusing of coherent
electron waves,\cite{Aidala07} and realizing chiral coherent quantum
circuits.\cite{Gabelli06-07}

One of the typical and significant issues in mesoscopic and
nanoscopic systems is time-dependent
transport.\cite{TangChu-Wu06,Tang01-Bylander05,Blumenthal07,%
Malshukov05-Kaun05-Pistolesi06,Szafran05,Okunishi07-Luan07}  A
microelectronic system driven by an external time-dependent
potential allows charge carriers to make coherent inelastic
scattering.  A number of time-dependent transport features have been
investigated such as time-dependent quasibound
states,\cite{TangChu-Wu06} nonadiabatic quantum charge
pumping,\cite{Tang01-Bylander05,Blumenthal07} current-driven
oscillations for nanomechanical
rectifiers,\cite{Malshukov05-Kaun05-Pistolesi06} and charged
particle motion in quantum rings.\cite{Szafran05,Okunishi07-Luan07}
Blumenthal \textit{et al.}\ demonstrated that the pumped current of
hundred picoamperes can be generated and is proportional to the
pumping frequency up to 3~GHz.\cite{Blumenthal07}  Szafran and
Peeters performed time-dependent simulation exploring the electron
wave packet trajectories in an open quantum ring by considering the
transport in the lowest subband and neglecting inelastic scattering
effects.\cite{Szafran05}

In the present work, our purpose is to elucidate how the embedded
quantum dots in a uniform perpendicular magnetic field affects the
transport characteristics of the electron wave packet in a broad
ballistic two-terminal quantum wire system.  By transforming the
embedded potential as well as the scattering wave function into a
momentum-coordinate mixed
representation,\cite{Gurvitz95-Gudmundsson05} we demonstrate that
the wave packet transmission probability and the conductance can be
obtained using the Lippmann-Schwinger method.\cite{DiVentra01} In
order to understand in detail, we shall consider embedded antidot
and double-dot systems in different magnetic fields for comparison.
In magnetic fields the propagating wave packet states are shifted to
the sample boundaries due to the Lorentz force.  Detailed
information on the embedded nanostructures represents a key to the
understanding of various features of the magnetotransport of a wave
packet.

The present paper is organized as follows. Section \ref{Model}
describes wave packet magnetotransport in a nanostructure embedded
quantum wire. In \sec{Results}, we examine wave packet propagation
of the quantum wire with embedded quantum dots and the robustness of
the resonance features in appropriate magnetic fields. Concluding
remarks and possible future directions are summarized in
\sec{Conclusion}.

%---------------------------------------------------------------------------
\section{Theoretical Model}\label{Model}

The system under investigation is a two-dimensional quantum wire
containing an embedded nanostructure penetrated by a perpendicular
magnetic field ${\bf B}= B {\bf \hat{z}}$. The quantum wire lies in
the $x$-$y$ plane, which is assumed to be confined in the $y$
direction and transport is in the $x$ direction. The wire stretches
into infinity in both directions while the embedded nanostructure is
contained in a scattering region of finite length in the middle of
the wire.  The Hamiltonian of system contains an unperturbed
Hamiltonian and a scattering potential describing the embedded
quantum dots, namely ${\cal H}={\cal H}_0 + V_{\mathrm{sc}}(x,y)$.
In this work, we shall explore the time-dependent transport
phenomena of embedded antidot and parallel double-dot systems.

\begin{figure}[htb]
     \includegraphics[width=0.86\linewidth]{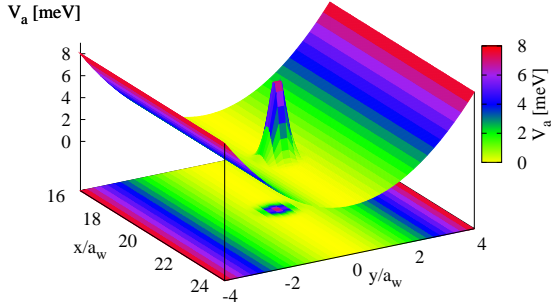}
     \caption{(color online). Schematic view of an antidot embedded in a
     two-terminal quantum wire. $V_{a0}=8$~meV,
     $\beta_a=10^{-2}$~nm$^{-2}$, and $x_a=20a_{\rm w}$.}
     \label{sysanti}
\end{figure}
The considered embedded antidot is defined in the middle of the
quantum wire as shown in \fig{sysanti}, and can be described by a
Gaussian-type potential
\begin{equation} \label{V-a}
     V_a(x,y) = V_{a0} \exp\left[ -\beta_a\left((x - x_a)^2 +y^2\right)
     \right],
\end{equation}

For performing numerical computation, the antidot related physical
parameters are selected as follows: potential height of the antidot
$V_{a0}=8$~meV, the potential broadening parameter
$\beta_a=10^{-2}$~nm$^{-2}$, and the longitudinal coordinate center
$x_a=20a_{\rm w}$ with $a_{\rm w}$ being the effective magnetic
length of the wire to be defined later.

\begin{figure}[htb]
     \includegraphics[width=0.86\linewidth]{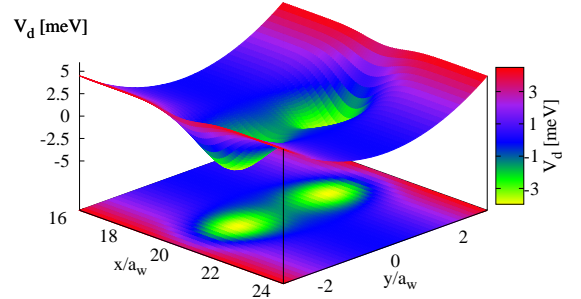}
     \caption{(color online). Schematic view of a parallel double-dot embedded in a
     two-terminal quantum wire.  $V_{d0}=-5$~meV, $\beta_d=10^{-3}$~nm$^{-2}$,
     $x_d=20a_{\rm w}$, and $y_d=1.5 a_{\rm w}$.} \label{sysDD}
\end{figure}

The parallel double-dot potential under consideration, shown in
\fig{sysDD}, is described by a number of Gaussian-type potentials
\begin{equation}\label{V-d}%
     V_d = V_{d0}  \exp\left[ -\beta_d (x-x_d)^2 \right]
     \sum_{\nu=\pm} \exp \left[ -\beta_d (y+\nu y_d)^2 \right]
\end{equation}
The strength of the coupling of the two parallel open quantum dots
is tunable by the separation parameter $y_d$ and the strength of the
magnetic field.  In our numerical calculation, we shall select the
strength of the double quantum dot $V_{d0}=-5$~meV, the broadening
parameter $\beta_d=10^{-3}$~nm$^{-2}$, the longitudinal center
$x_d=20 a_{\rm w}$, and the transverse off center parameter
$y_d=1.5a_{\rm w}$ such that the two dots are separated by the
distance of $2y_d$.

In the Landau gauge for the vector potential, the unperturbed
Hamiltonian can be written as
\begin{equation}\label{H-0}
      {\cal H}_0=\frac{\hbar^2}{2m^*}\left[-i{\bf \nabla}-\frac{eB}{\hbar
      c}y{\bf \hat{x}} \right]^2 + V_{\rm conf}(y),
\end{equation}
where $-e$ and $m^*$ are the charge and the effective mass of an
electron, respectively.  The confining potential $V_{\rm
conf}(y)=\frac{1}{2}m^*\Omega_0^2 y^2$ is assumed to be parabolic.
Using a mixed momentum-coordinate
representation\cite{Gurvitz95-Gudmundsson05} and making Fourier
transform in time, the scattering wave function
\begin{equation}\label{Psi-x}
      \Psi(x,y,t)=\int_{-\infty}^{\infty}\frac{d\omega}{\sqrt{2\pi}}
      \int_{-\infty}^{\infty}\frac{dp}{\sqrt{2\pi}}
      e^{i(px-\omega t)}\widetilde{\Psi}(p,y,\omega)
\end{equation}
can be separated into the coefficient functions $\varphi_n$ and the
shifted harmonic-oscillator-type eigenfunctions $\Phi_n$ for the
wire, namely
$\widetilde{\Psi}(p,y,\omega)=\sum_n\varphi_n(p,\omega)\Phi_n(y-y_p)$
where the shifting center $y_p$ = $p a_{\rm w}^2\omega_c/\Omega_{\rm
w}$ is momentum dependent.  In the absence of magnetic field, this
shifting center is identically zero.  The effective magnetic length
of the wire $a_{\rm w}=\hbar/m^*\Omega_{\rm w}$ is related to the
effective cyclotron frequency $\Omega_{\rm
w}=\sqrt{\Omega_0^2+\omega_c^2}$ with $\omega_c=eB/(m^*c)$ being the
two-dimensional cyclotron frequency. In the presence of a magnetic
field, the quantized electron energy away from the scattering region
is given by\cite{Berggren89-Tang-Vidar}
\begin{equation}
E_n(p) = E_n^0 + \frac{U_{\rm w}}{2}\left( p a_{\rm w} \right)^2  \:
,
\end{equation}
where $E_n^0$ = $(n+1/2)E_{\rm w}$ are the transverse subband energy
levels, and the second term denotes the kinetic energy with $U_{\rm
w}$ = $(\hbar\Omega_0)^2/E_{\rm w}$ and $E_{\rm w}$ =
$\hbar\Omega_{\rm w}$.

To obtain the coefficient functions $\varphi_n$, one defines the
momentum-coordinate space potential $V(p-q,y)$, which is a Fourier
transform of the scattering potential $V_{\rm sc}(x,y)$.  The
overlap integral in the momentum space can thus be expressed as
\begin{eqnarray}
       U_{nn'}(p,q) U_{\rm w} &=& \int_{-\infty}^{\infty}dy\Phi_{n'}^*(y-y_q)\nonumber \\
       &&\times V(p-q,y)\Phi_n(y-y_p),
\end{eqnarray}
where $U_{nn'}$ is a dimensionless quantity.  In the asymptotic
region away from the scattering region, the unperturbed Green
function can be expressed of the form $G_n(p,\omega) =
[(k_n(\omega)a_{\rm w})^2-(pa_{\rm w})^2]^{-1}$, where the
dimensionless wave vector $k_n(\omega)a_{\rm w}$ = $[(\hbar \omega
-E_n^0)/U_{\rm w}]^{1/2}$ describes the dispersion relation in the
asymptotic regions. After some algebra, one can obtain the
Lippmann-Schwinger equation in the momentum space
\begin{eqnarray}
      \varphi_n(p,\omega)&=&\varphi_n^0(p,\omega)
      +G_n(p,\omega)\int_{-\infty}^{\infty}\frac{dq a_{\rm w}}{\sqrt{2\pi}}\nonumber\\
      && \times U_{nn'}(p,q)\varphi_{n'}(q),
\end{eqnarray}
where $\varphi_n^0(p,\omega)=2\pi g_n(p)\delta [ \omega -
E_n(p)/\hbar ]$ is the coefficient function of the asymptotic
regions.  Therein, the envelope function of the incident wave packet
$g_n(p)$ is assumed to contain only positive $p$ values such that
the wave packet is injected in the ${\bf x}$ direction. For a given
subband $n$, the explicit form of the coefficient function can be
expressed in terms of the $T$-matrix
\begin{eqnarray}
      \varphi_n(p,\omega)
      &=& \varphi^0_n(p,\omega) +
      G_n(p,\omega) \sum_{n'}\int_{-\infty}^{\infty} \frac{dq a_{\rm w}}{\sqrt{2\pi}} \nonumber \\
      &&\times T_{nn'}(p,q,\omega)\varphi_{n'}^0(q,\omega),
\end{eqnarray}
where the $T$-matrix is a solution of the integral equation
\begin{eqnarray}
      T_{nn'}(p,q,\omega) &=& U_{nn'}(p,q)
      + \sum_m\int_{-\infty}^{\infty}\frac{dk a_{\rm w}}{\sqrt{2\pi}}U_{nm}(p,k) \nonumber \\
      &&\times G_m(k,\omega)T_{mn'}(k,q,\omega).
\end{eqnarray}
Solving this integral equation for the $T$-matrix, one can obtain
the coefficient functions $\varphi_n$ for the scattering region and
construct the total wave function $\Psi(x,y,t)$ =
$\Psi_0(x,y,t)+\Psi_{\mathrm{sc}}(x,y,t)$ containing an asymptotic
part
\begin{eqnarray}
      \Psi_0(x,y,t) &=& \sum_n \int_{-\infty}^{\infty} dp
g_n(p) \Phi_n(y-y_p) \nonumber \\
&&\times \exp \left\{ i\left[ px - E_n(p)t/\hbar \right]
      \right\}
\end{eqnarray}
with $g_n(p)$ = $\delta_{nn'}\exp[-\gamma (p-p_0)^2]$ being the
envelop function of the wave packet in momentum space, and a
scattering part
\begin{eqnarray}
      \Psi_{\rm sc}(x,y,t) &=& \sum_n\int_{E_n^0/\hbar}^{\infty}d\omega e^{-i\omega
t}\frac{\Omega_{\rm w} g_n[k_n(\omega)]}{\Omega_0^2 \vert
k_n(\omega)a_{\rm w} \vert}\nonumber \\
&&\times \sum_{n'}\int_{-\infty}^{\infty}
      \frac{dpa_{\rm w}}{\sqrt{2\pi}}G_{n'}(p,\omega)\exp(ipx)\nonumber \\
      &&\times T_{n'n}(p,k_n(\omega))
      \Phi_n(y-y_p)\: .
\end{eqnarray}

The transmission amplitude through the embedded quantum dot system
for an electron with energy $E=\hbar\omega$ entering the scattering
region from the channel $n$ in the left lead and leaving it via
channel $m$ in the right lead that can be expressed in terms of the
$T$-matrix:
\begin{equation}\label{t-nm}
    {\bf t}_{nm}(\omega)=\delta_{nm}-\frac{i}{2k_m(\omega)}\frac{2m}{\hbar^2}T_{nm}
    \left[ k_n(\omega),k_m(\omega) \right] \:.
\end{equation}
The conductance, according to the framework of multichannel
Landauer-B{\"u}ttiker formalism,\cite{Landauer-Buttiker} is written
as
\begin{equation}
      G = G_0 \mbox{Tr}[{\bf t}_{nm}^\dagger(\omega){\bf t}_{nm}(\omega)]\: ,
\label{G}
\end{equation}
where $G_0 = 2e^2/h$ is the universal conductance quantum and  ${\bf
t}_{nm}$ is evaluated at Fermi energy.  All the incident and
scattered propagating modes have to be taken into account.

%-----------------------------------------------------------------------------------------------
\section{Numerical Results}\label{Results}

To investigate the magnetotransport properties of wave packet
propagation in a nanostructure system embedded in a broad wire under
a perpendicular magnetic field, we select the confinement parameter
$\hbar\Omega_0=1$~meV.   In our numerical calculation, the magnetic
field strengths are selected as $B=0.5$ and $1.0$~T with
corresponding effective magnetic lengths $a_{\rm w}=29.3$ and
$23.9$~nm, respectively.  We assume that the quantum wire is
fabricated in a high-mobility GaAs-Al$_x$Ga$_{1-x}$As
heterostructure such that the effective Rydberg energy
$E_{\mathrm{Ryd}}=5.92$~meV and the Bohr radius $a_{\rm B}=9.79$~nm.
Below we shall explore the dynamic motion of the electron wave
packet in a quantum wire under an applied perpendicular magnetic
field with either an embedded antidot or an embedded double open
quantum dot.

\subsection{Embedded Antidot}\label{EA}

Earlier work considering magnetotransport in an antidot was carried
out by assuming that magnetic field is so strong that only the
lowest Landau level is occupied.\cite{Sanchez04}  The antidot can be
formed by producing a potential hill with
gates,\cite{Kirczenow94-Kataoka02} and behaves effectively like an
artificial quantum impurity.  It is thus warranted to devote further
effort in developing numerical techniques in order to analyze the
behavior of the electron wave packet propagation in a quantum wire
with an embedded antidot in a tunable magnetic field.

\begin{figure}[htb]
     \includegraphics[width=0.86\linewidth]{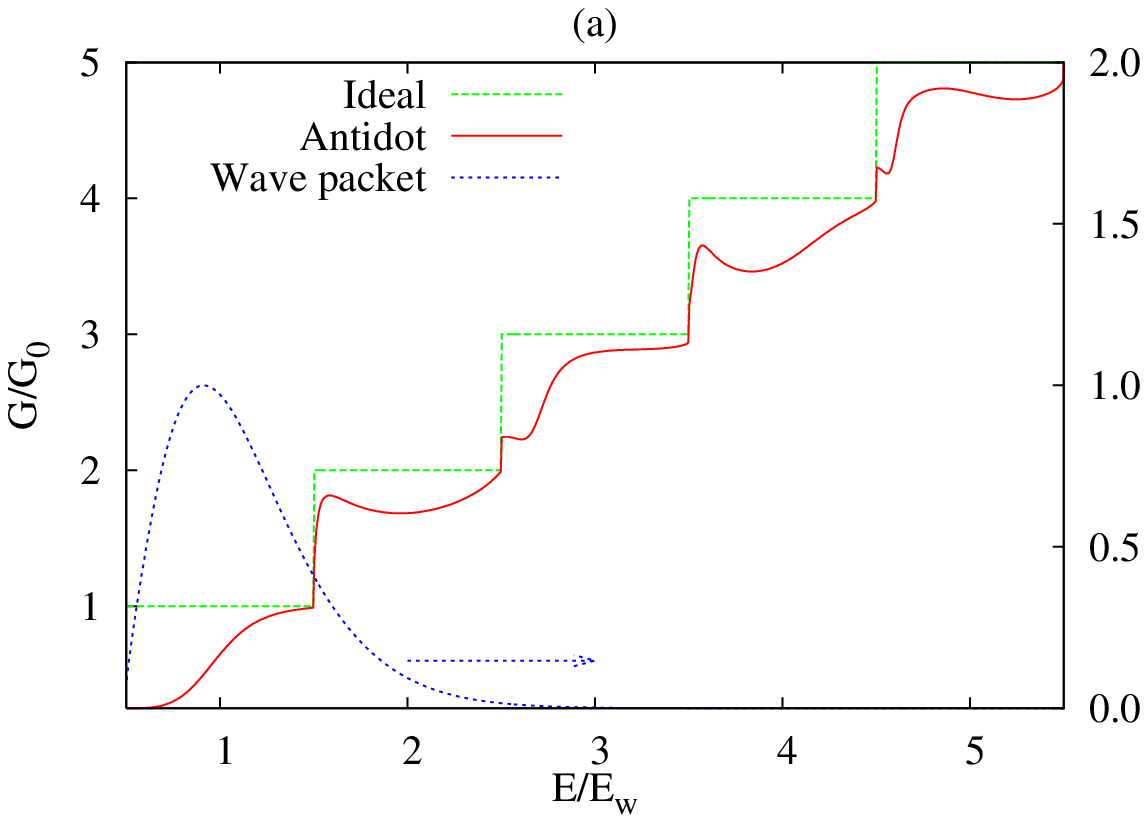}
     \includegraphics[width=0.86\linewidth]{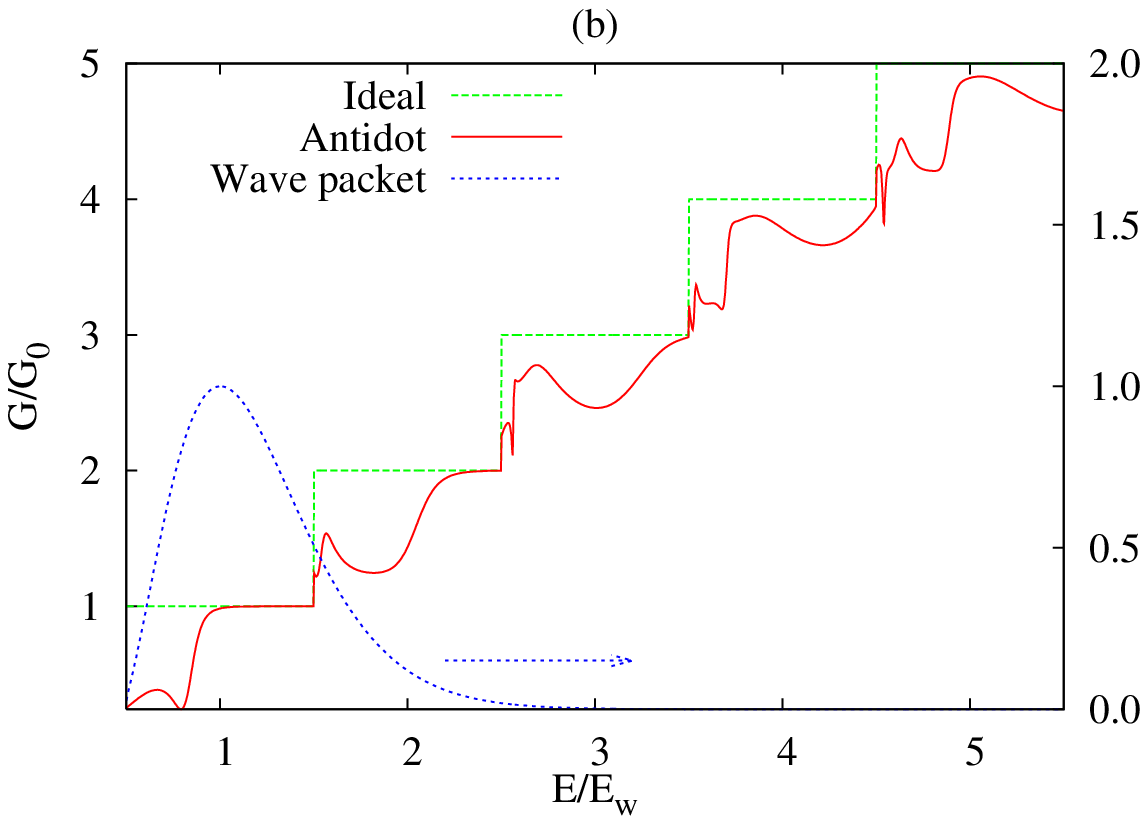}
     \caption{(color online). Energy dependence of the conductance
     in an ideal wire (dashed green),
     the conductance in an antidot embedded wire (solid red),
     and the envelop function of the wave packet in momentum space (dotted blue)
     for the cases of (a) $B=0.5$ with wave packet parameters
     $p_0=1.2a_{\rm w}^{-1}$ and $\gamma=2.0a_{\rm w}^2$; and (b)
     $B=1.0$ T with wave packet parameters
     $p_0=2.0a_{\rm w}^{-1}$ and $\gamma=1.0a_{\rm w}^2$.
     The other parameters are the same as \fig{sysanti}.}
     \label{antidot-G}
\end{figure}

Since the effective magnetic length $a_{\rm w}$ is a function of
magnetic field, we thus select the envelop parameters of the wave
packet in the momentum space as $p_0=1.2a_{\rm w}^{-1}$ and
$\gamma=2.0a_{\rm w}^2$ for $B=0.5$~T, and $p_0=2.0a_{\rm w}^{-1}$
and $\gamma=1.0a_{\rm w}^2$ for $B=1.0$~T, such that the wave
packets are of similar shapes in momentum space as shown by the
dotted blue curve in \fig{antidot-G}. The incident wave packet is
selected to have contributions from the lowest subband for clarity.
The initial electron envelop function at $t=0$ is a Gaussian wave
packet in the momentum space with width $\Delta p_{\rm
in}=1/\sqrt{\gamma}$ such that the probability density of the wave
packet in the momentum space is reduced by a factor of $1/\sqrt{e}$.

The energy dependence of the conductance for the traveling wave
packet in an ideal wire and an antidot embedded wire are depicted in
\fig{antidot-G} by the dashed green and the solid red curves,
respectively. The general feature in \fig{antidot-G} is that the
conductance is generally suppressed in the low kinetic energy regime
but approaches the conductance of the ideal wire in the high kinetic
energy regime. This is because the electron waves with lower kinetic
energy are easier to be backscattered by the embedded antidot.  For
the case of $B=0.5$~T, the conductance manifests a smooth transition
region in the lowest subband ($n=0$).  Low kinetic energy blocking
phenomenon is significant at the third and the fifth conductance
plateaus. However, in the second and the fourth plateaus, the
Lorentz force pushes transversely the electron wave packet with
mediate kinetic energy, and hence suppresses slightly the
conductance plateaus.

For the case of $B=1.0$~T, the conductance in the lowest subband
region exhibits clear transition between the backward and the
forward propagating energy regimes at around $E\simeq E_{\rm w}$. A
dip structure is clearly found at $E\simeq 0.79E_{\rm w}$ that
corresponds to a short-lived quasibound state with negative binding
energy.  Such a dip structure becomes broader valley structure at
higher subbands shifted slightly to the higher energy.  This
broadening indicates a shorter dwell time of the localized state at
higher subbands. Sharp dip structures at $E=2.56$, 3.52, and
4.54$E_{\rm w}$ demonstrate the formation of quasibound states with
negative binding energy.\cite{Vidar05}

\begin{figure*}[htb]
     \includegraphics[width=0.35\linewidth]{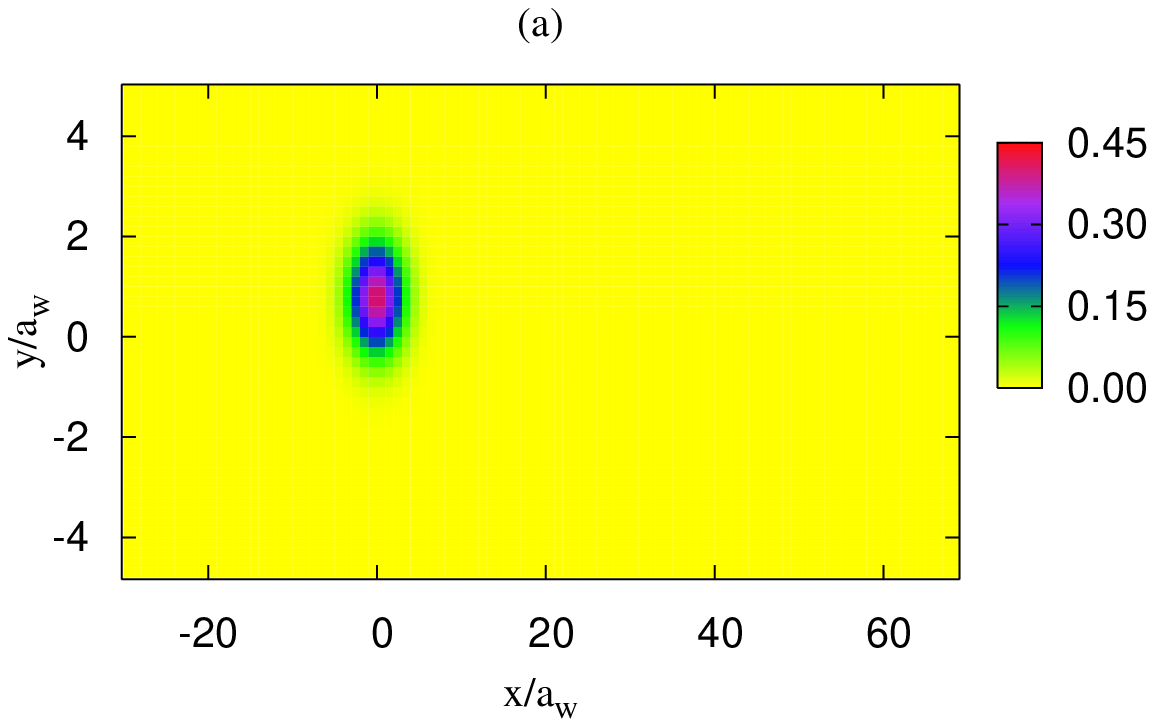}
     \includegraphics[width=0.35\linewidth]{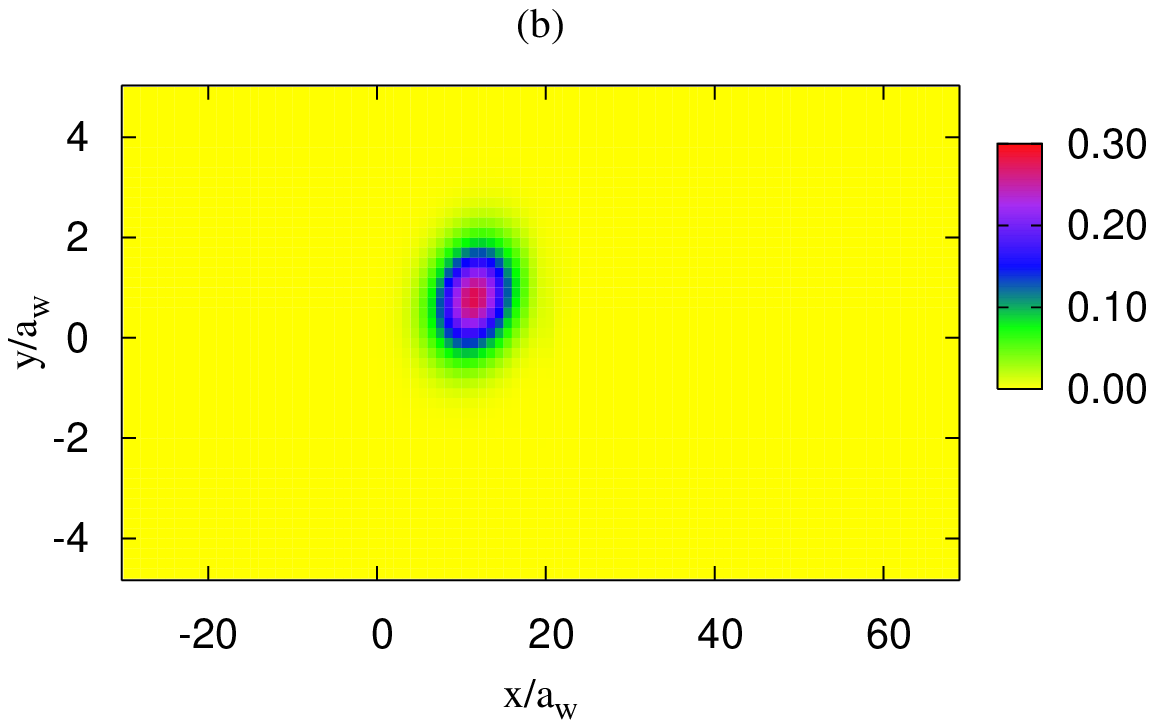}
     \includegraphics[width=0.35\linewidth]{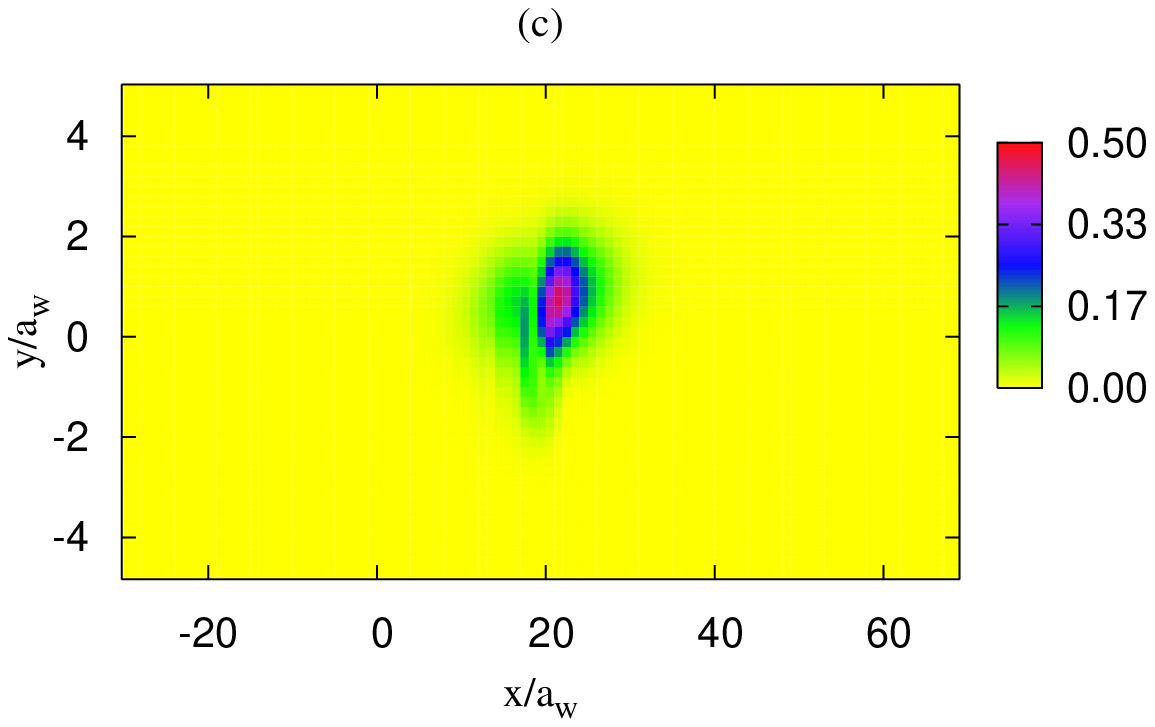}
     \includegraphics[width=0.35\linewidth]{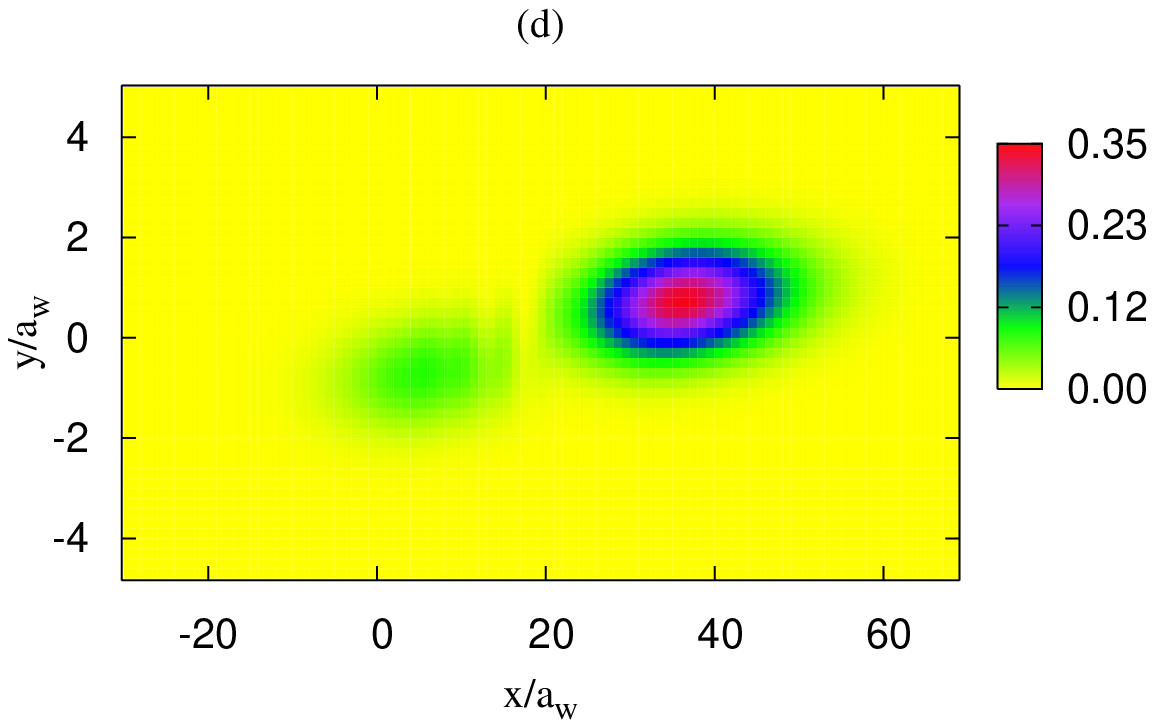}
     \caption{(color online). Propagation of the electron wave packet traveling through
     an embedded antidot for the case of $B=0.5$~T at time $t=$ (a) 0; (b) 8; (c) 15;
     and (d) 28~ps.  The other parameters are the same as \fig{sysanti}.}
     \label{Antidot-05T}
\end{figure*}
In \fig{Antidot-05T}, we show the time evolution of the wave packet
traveling through the quantum wire with an embedded antidot for the
case of $B=0.5$~T.  Before the electron wave packet arrives at the
scattering region, the wave packet center is shifted slightly in
transverse direction to $y$$\approx$0.8$a_{\rm w}$ due to the
Lorentz force induced by the penetrating magnetic field as shown in
\fig{Antidot-05T}(a) for $t=0$~ps.  To estimate the longitudinal
width of the incident Gaussian wave packet, it is convenient to
define a Gaussian function in the real space $f(x)=e^{-x^2/\gamma}$
with width $\Delta x$ such that $x$ varies from 0 to $\pm \Delta
x/2$, and $f(x)$ is reduced by a factor of $e^{-1/2}$.  By this
definition, one can estimate the width of the incident wave packet
at $t=0$ being $\Delta x_{\rm in}=2\sqrt{\gamma}a_{\rm w}$ to obtain
$\Delta x_{\rm in}\Delta p_{\rm in}=2$ which is compatible with the
Heisenberg uncertainty relation.

Figure \ref{Antidot-05T}(b) shows the time evolution of the wave
packet at $t=8$~ps, it is found that the trajectory of the electron
waves with higher kinetic energy is closer to the wire edge due to
the magnetic field. Since the higher kinetic energy electron waves
contain larger group velocity, the shape of the wave packet is
skewed.  The electron wave packet is then scattered by the antidot
as shown in \fig{Antidot-05T}(c).  The electron waves with higher
kinetic energy can pass through the antidot but the lower kinetic
energy part of the wave packet is reflected.  Due to the Lorentz
force, the backscattered wave packet is turned around to lower part
of the quantum wire with wave packet center at $y=-0.6a_{\rm w}$
[see \fig{Antidot-05T}(c)]. For longer evolution time $t=28$~ps, the
scattered wave packets become broader and then leave the scattering
region, as is shown in \fig{Antidot-05T}(d).

\begin{figure*}[htb]
     \includegraphics[width=0.35\linewidth]{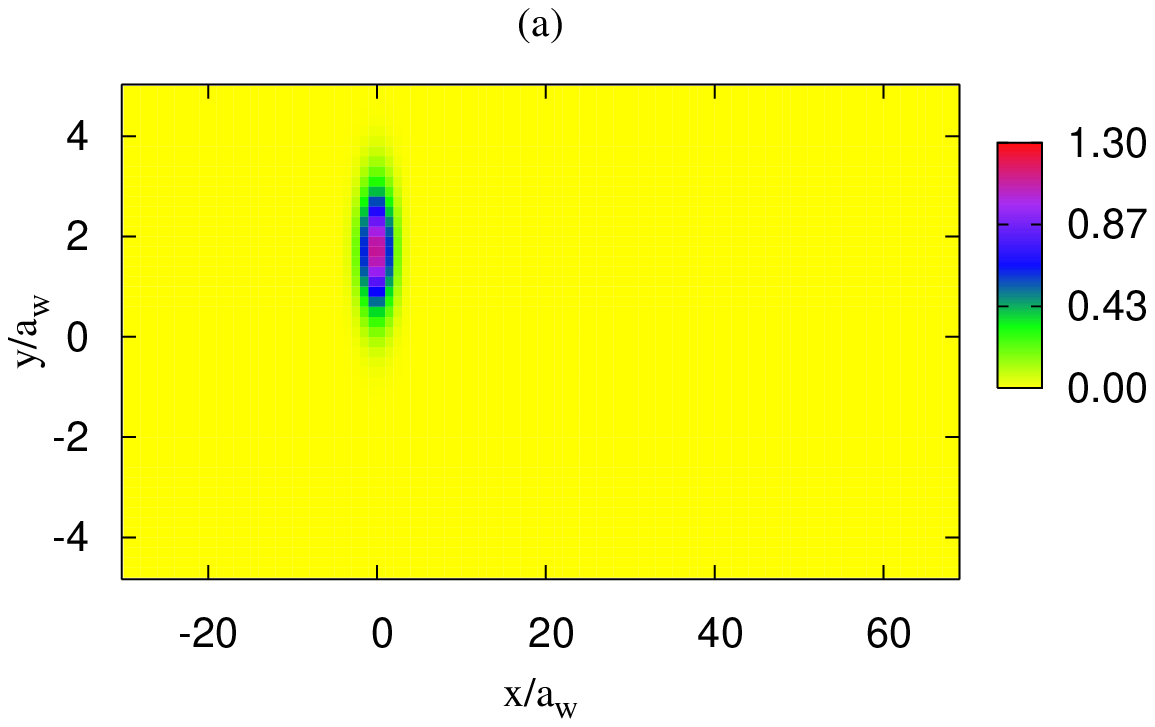}
     \includegraphics[width=0.35\linewidth]{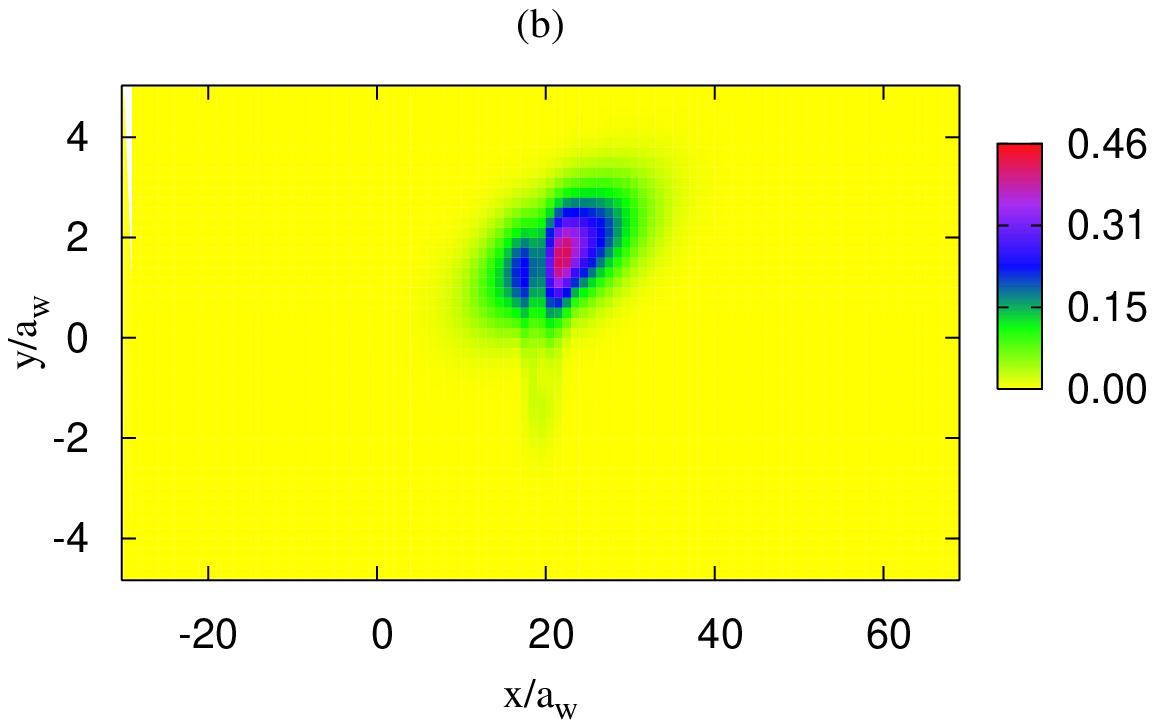}
     \includegraphics[width=0.35\linewidth]{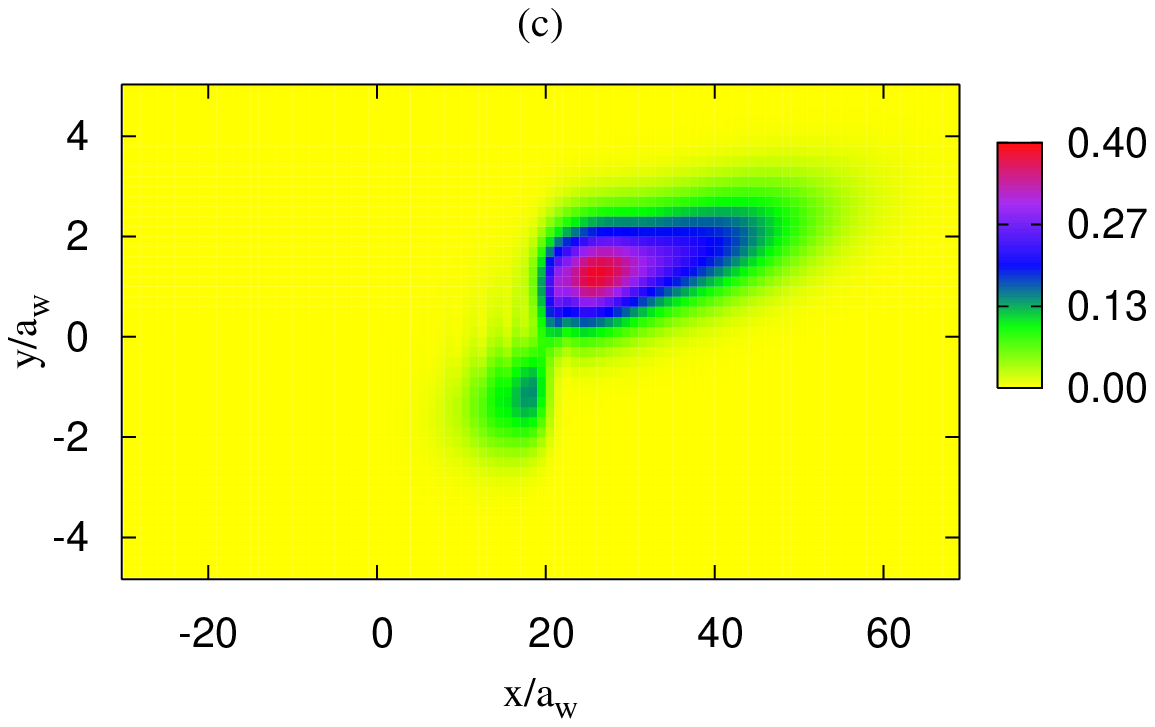}
     \includegraphics[width=0.35\linewidth]{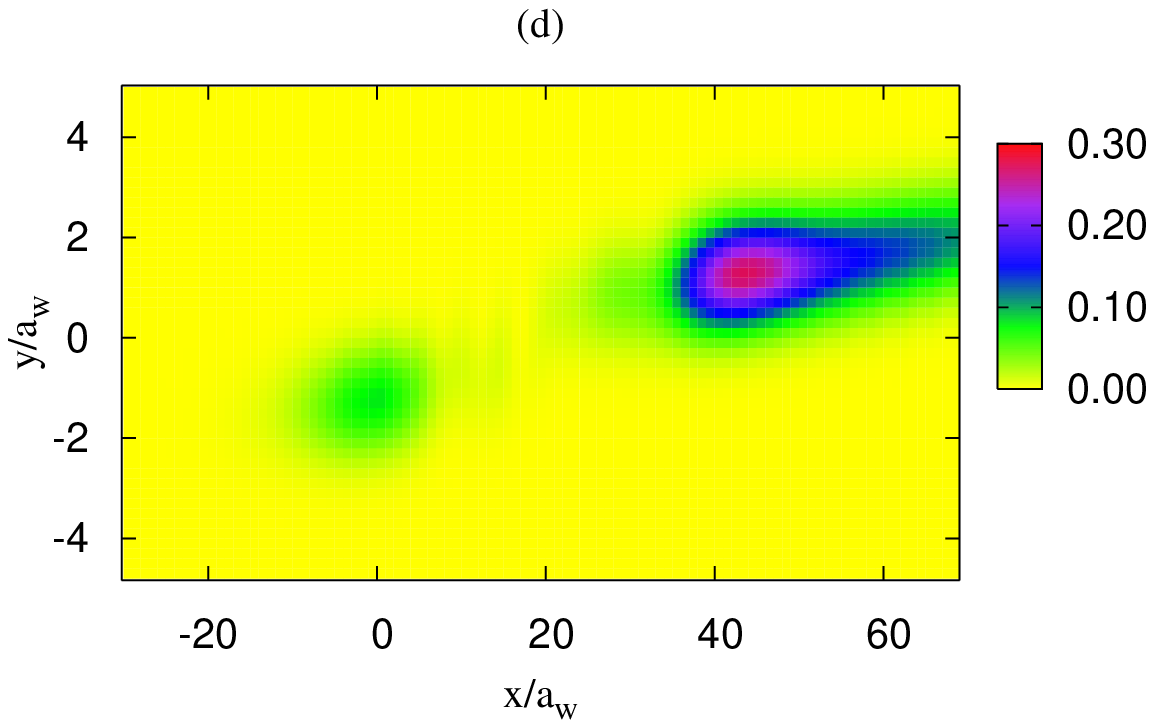}
     \caption{(color online). Propagation of the electron wave packet traveling through
     an embedded antidot for the case of $B=1.0$~T at the time $t=$ (a) 0; (b) 15; (c) 25;
     and (d) 40~ps.  The other parameters are the same as \fig{sysanti}.}
     \label{Antidot_1T}
\end{figure*}
In \fig{Antidot_1T}, we show the time evolution of the wave packet
traveling through the quantum wire with an embedded antidot for the
case of $B=1.0$~T. Before the electron wave packet arrives at the
scattering region, the wave packet has a shifting center at
$y$$\approx$1.8$a_{\rm w}$ due to the Lorentz force as shown in
\fig{Antidot_1T}(a) for zero picoseconds. The wave packet is narrow
in the $x$ direction, that is, $\Delta x_{\rm in}=2a_{\rm w}$.  The
electron wave packet is scattered by the antidot at $t=15$~ps, as is
shown in \fig{Antidot_1T}(a). The part with higher kinetic energy
can pass through the antidot but the part with lower kinetic energy
is predominately reflected.  Due to the magnetic field induced
Lorentz force, the backscattered wave packet is turned around to
lower part of the quantum wire with wave packet center at
$y=-1.2a_{\rm w}$ at $t=25$~ps (see \fig{Antidot_1T}(c)). For longer
time $t=40$~ps, the scattered wave packets are getting broader---the
reflected wave packet has distribution length $\Delta x_{\rm
ref}$$\approx$$10a_{\rm w}$ and the transmitted wave packet has even
broader distribution length $\Delta x_{\rm tran}$$\approx$$20a_{\rm
w}$, as is shown in \fig{Antidot_1T}(d).  The spreading of a wave
packet is a quantum diffusion phenomenon, which was utilized for
possible application in a quantum kicked rotor system.\cite{Zhong01}

In comparison to the incident wave packet in an applied magnetic
field $B=1.0$~T, the incident wave packet in $B=0.5$~T has a wider
longitudinal profile and a wave packet center closer to the middle
of the quantum wire.  It turns out that the group velocity of the
wave packet for the case of $B=0.5$~T is greater than that of
$B=1.0$~T. We would like to mention in passing that for a wave
packet with momentum envelop function covering more subbands, the
electron wave packet tends to form transversely skipping-like
trajectories in both the forward and backward scattered wave packets
due to the mode mixing interference between different subbands.

\subsection{Embedded Parallel Double Dot}\label{EPDD}

Electronic transport through coupled quantum nanostructures is of
fundamental interest for the understanding of coherent resonant and
superposition states.  By coupling two quantum dots in
series\cite{vdWiel03} or in parallel,\cite{Chen04} a double quantum
dot is formed. Quantum transport through such a double-dot system
has attracted considerable attention due to its versatility for
various applications such as probing
entanglement,\cite{Loss00-Schomerus07} detecting microwave
manipulation of a single electron,\cite{Petta04} analyzing dephasing
rate,\cite{Elhassan05} studying nonadiabatic transport under
irradiation,\cite{Naber06} and readout of the coherent superposition
of trajectories.\cite{Jordan06}  The interdot coupling strength can
be experimentally varied using gate electrodes.\cite{Chen04,Craig04}

Earlier works considering electronic transport in double quantum dot
systems were carried out using the Anderson-type hopping model by
assuming that the system is isolated with weak coupling to the
leads.\cite{Kiselev03-Hartmann04-Zarand06}  Our previous work has
devoted effort in developing numerical computation of
magnetotransport in a transversely hill-separated parallel double
open quantum dot system with strong coupling to the
leads.\cite{Tang05} It is thus appropriate to analyze the
propagation behavior of the electron wave packet in a quantum wire
with an embedded parallel double open quantum dot to get better
insight into the dynamical properties.
\begin{figure}[htb]
     \includegraphics[width=0.86\linewidth]{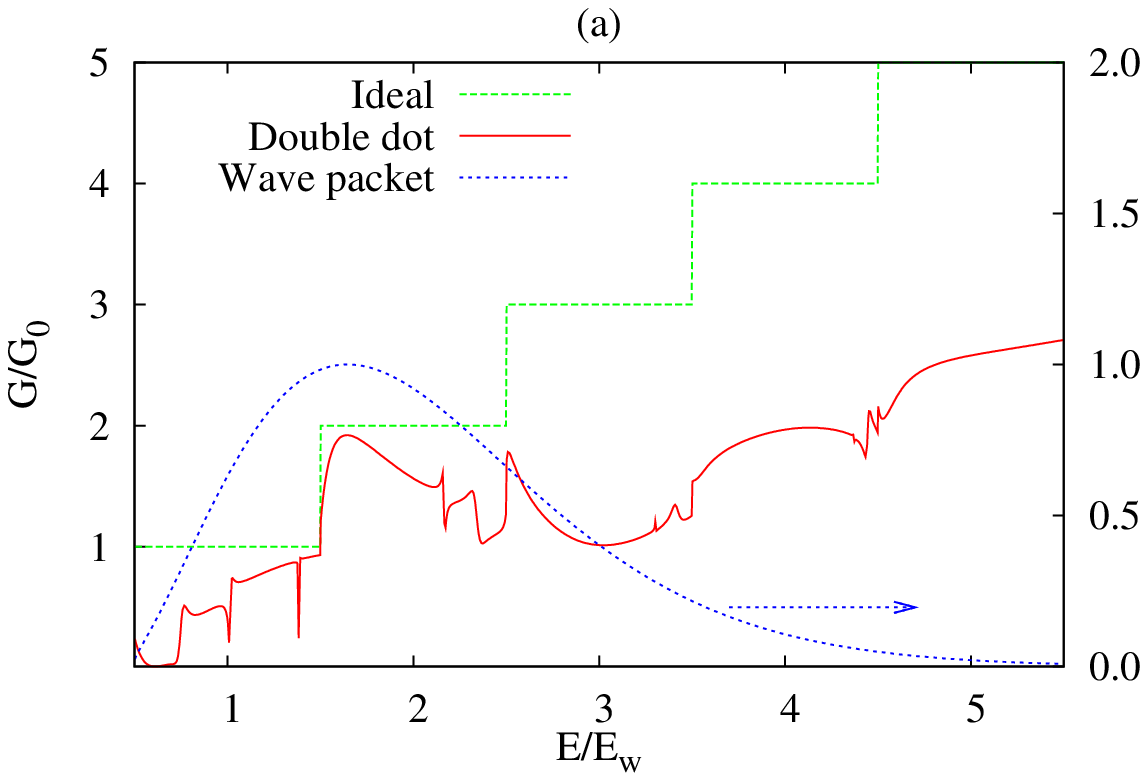}
     \includegraphics[width=0.86\linewidth]{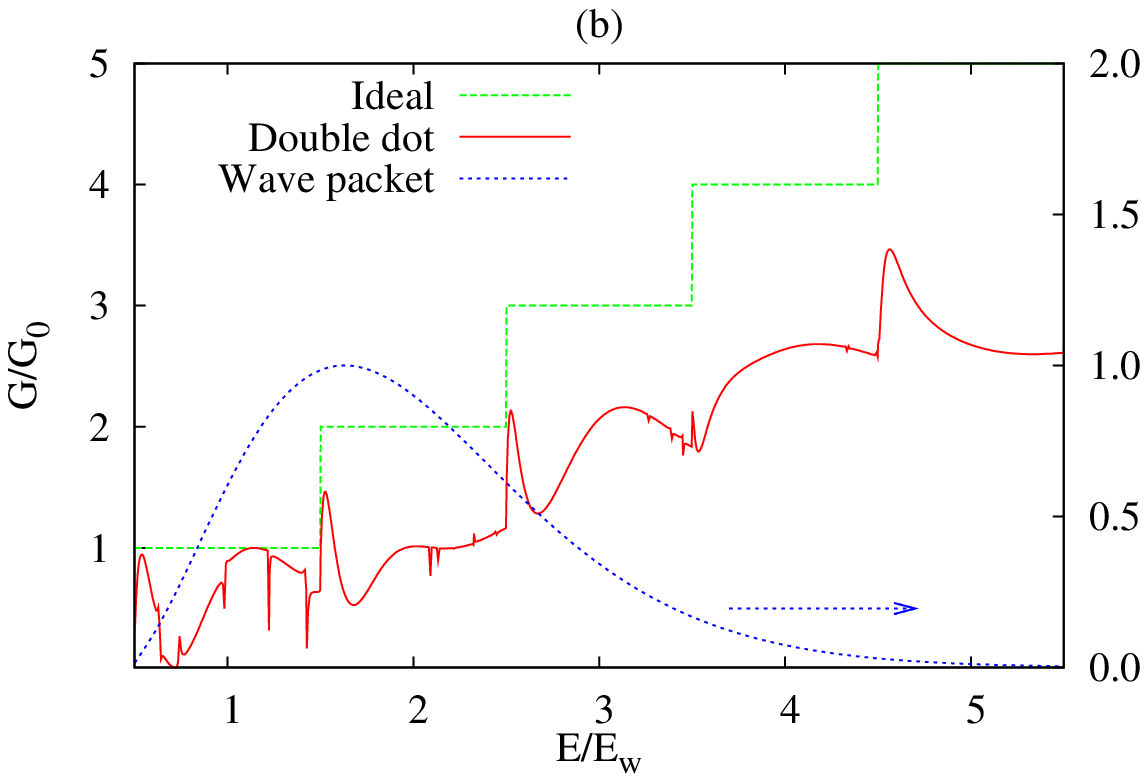}
     \caption{(color online).
     Energy dependence of the conductance in an ideal wire (dashed green),
     the conductance in a double-dot embedded quantum wire (solid red),
     and the envelop function of the wave packet in momentum space (dotted blue)
     for the cases of (a) $B=0.5$ with wave packet parameters
     $p_0=1.0a_{\rm w}^{-1}$ and $\gamma=2.0a_{\rm w}^2$; and (b)
     $B=1.0$ T with wave packet parameters
     $p_0=3.0a_{\rm w}^{-1}$ and $\gamma=0.5a_{\rm w}^2$.
     The other parameters are the same as \fig{sysDD}.}
     \label{G-DD}
\end{figure}

In \fig{G-DD}, we show the energy dependence of the conductance in
an ideal wire (dashed green), the conductance in a double-dot
embedded quantum wire (solid red), and the envelop function of the
wave packet in momentum space (dotted blue) for the cases of (a)
$B=0.5$, and (b) $B=1.0$~T.  We select the envelop parameters of the
wave packet in the momentum space as packet center $p_0=1.0a_{\rm
w}^{-1}$ and $\gamma=2.0a_{\rm w}^2$ for $B=0.5$~T, and
$p_0=3.0a_{\rm w}^{-1}$ and $\gamma=0.5a_{\rm w}^2$ for $B=1.0$~T,
such that the wave packets are of similar shapes in momentum space.

For the case of $B=0.5$~T, shown in \fig{G-DD}(a), we see that a
perfect conductance gap formed at the kinetic energy regime
($0.53<E/E_{\rm w}<0.73$) of the first subband.  This fact indicates
that the embedded double-dot system may be applicable as a quantum
switch.  The conductance gap is formed due to the cyclotron motion
of electron wave between the two parallel dots.  Fano line-shapes in
conductance at energies $E=1.01$ and 2.37$E_{\rm w}$ manifest the
quantum interference feature of the wave packet between the part
forming quasibound states inside the double-dot system and the part
with straight transmission.  Furthermore, a sharp dip structure in
conductance at energy $E=1.38E_{\rm w}$ indicates the forming of a
quasibound state below the second subband threshold in the lead. The
transport properties of electron waves in the second subband is very
different to that in the first subband. The overall feature is that
the low kinetic energy electron exhibits higher conductance. The
strong suppression in conductance at higher subband implies the
better inter-dot coupling enhancing backscattering.

Figure \ref{G-DD}(b) shows the energy dependence of conductance for
the case of $B=1.0$~T.  The gap feature in conductance at the low
kinetic energy of the first subband is narrower than that induced by
the magnetic field $B=0.5$~T.  The dip structure in conductance
related to the formation of quasibound state at around the energy
$E=E_{\rm w}$ is almost the same as the case of $B=0.5$~T, but the
dip structure at higher energy in the first subband is shifted
toward the lower energy. A new clear sharp dip structure is formed
just below the threshold of the second subband. Since the wave
function of the electrons occupying the second subband in $B=1.0$~T
fits the geometry of the double-dot system, the electron wave thus
favors to turn around through the double-dot, and then the
conductance is strongly suppressed.  The conductance at energies
higher than the third subband threshold for the case of $B=1.0$~T is
a little higher than that of $B=0.5$~T.

\begin{figure*}[htb]
     \includegraphics[width=0.35\linewidth]{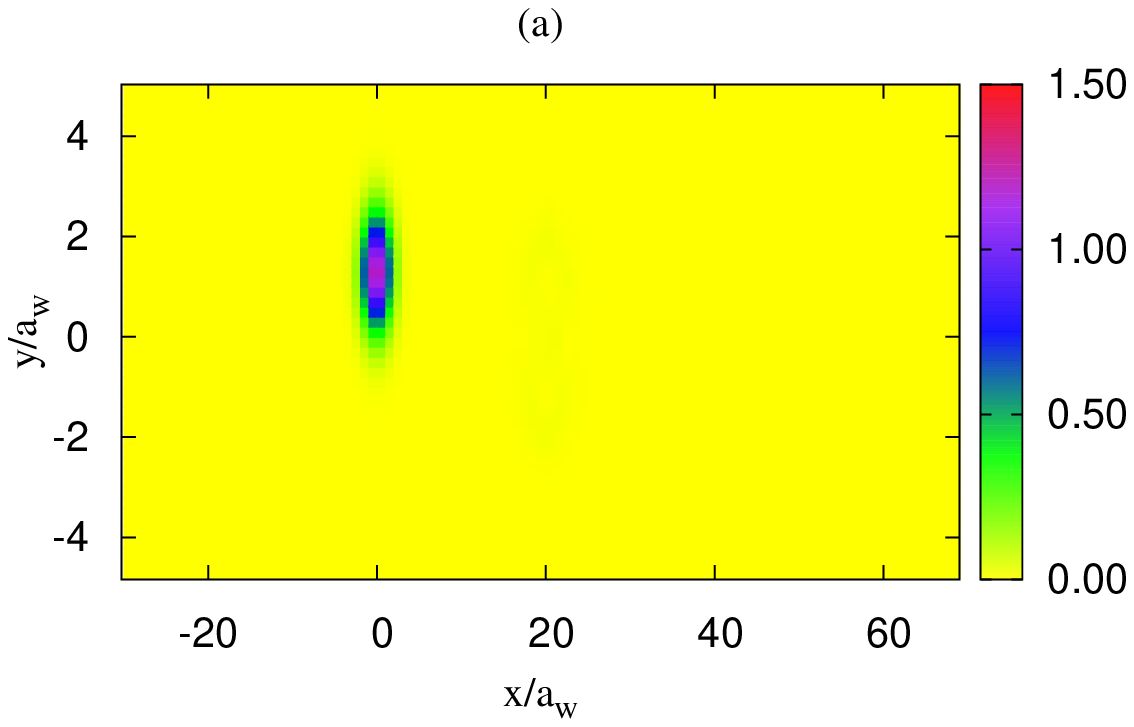}
     \includegraphics[width=0.35\linewidth]{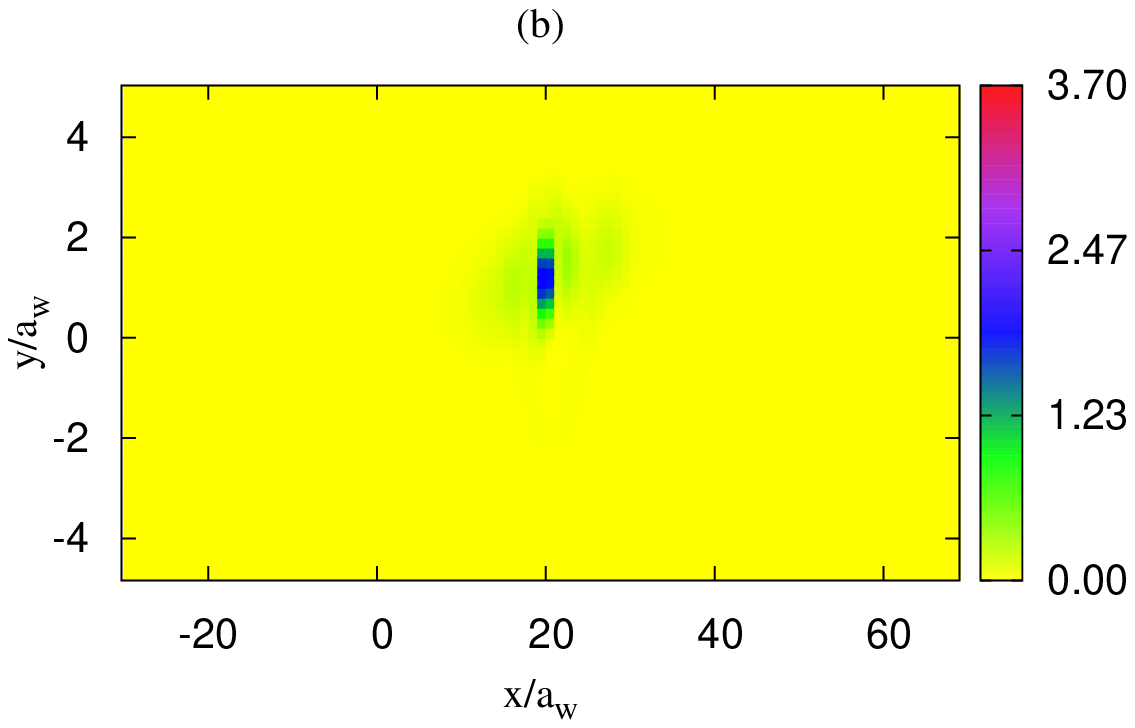}
     \includegraphics[width=0.35\linewidth]{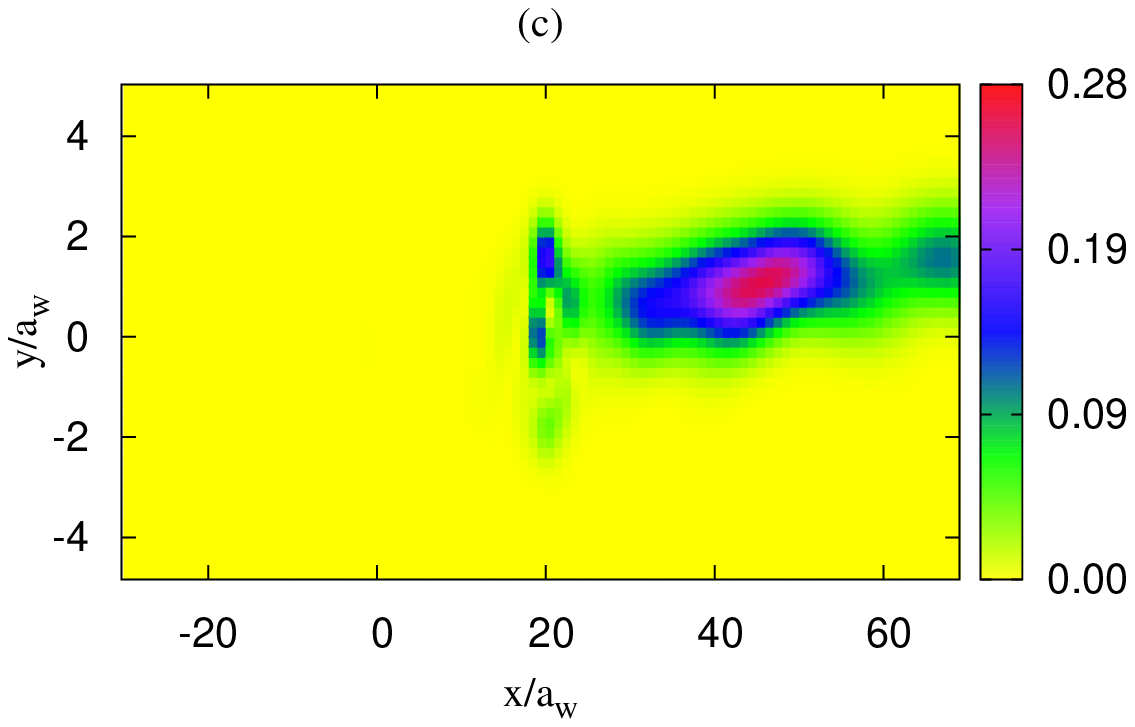}
     \includegraphics[width=0.35\linewidth]{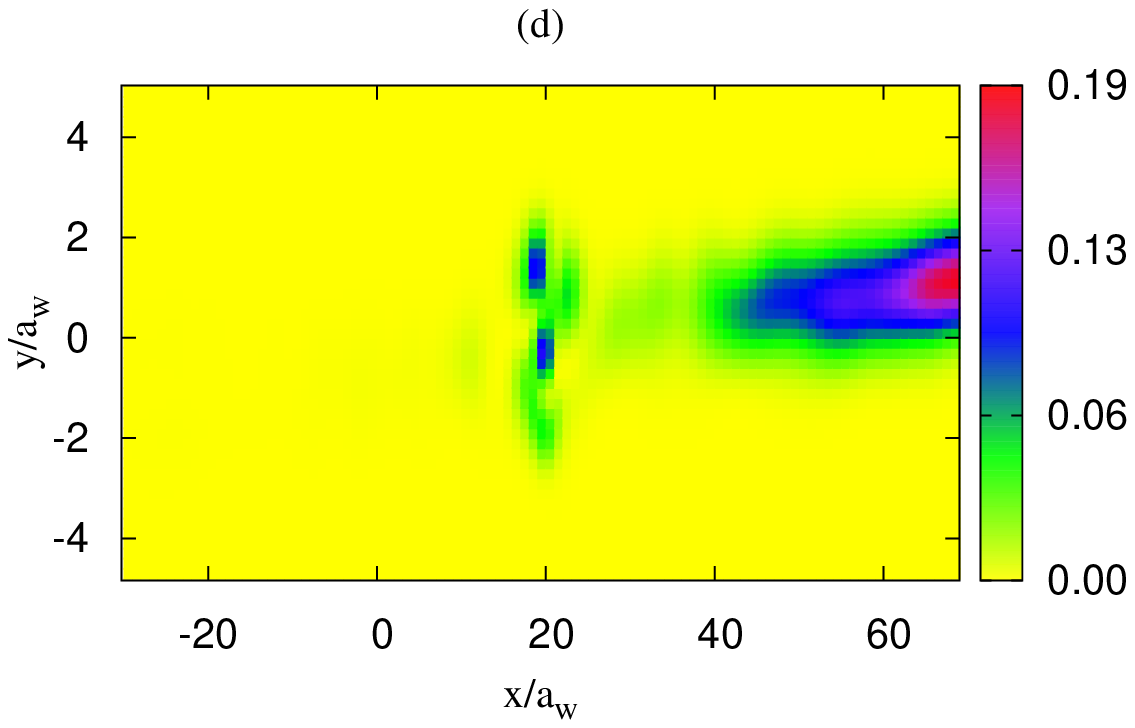}
     \caption{(color online). Propagation of the electron wave packet
     traveling through an embedded parallel double-dot system for
     the case of $B=0.5$~T at the time $t=$ (a) 0; (b) 9; (c) 25; and (d) 38~ps.
     The other parameters are the same as \fig{sysDD}.}
     \label{DD-05T}
\end{figure*}
Figure \ref{DD-05T} demonstrates the snapshots of the electron wave
packet propagation through an embedded parallel double-dot system
for the case of $B=0.5$~T at the time $t=$ (a) 0; (b) 9; (c) 25; and
(d) 38~ps.  At time $t=$0~ps, we see that the incident wave packet
has a compact longitudinal distribution $\Delta x_{\rm
in}=2\sqrt{2}a_{\rm w}$ with height 1.5.  In \fig{DD-05T}(b), at
$t=9$~ps, the electron wave packet arrives at the upper open dot and
form a clear quasibound state with packet height 2.0.  At this
moment, both the backward reflection and the forward transmission
are blocked by the double-dot system.

During the time evolution $0<t<9$~ps, the higher energy part of the
electron wave packet is closer to the upper boundary and traveling
faster than the lower energy part of the wave packet. This makes the
electron wave packet to skew with a clockwise rotation. Before being
scattered by the double-dot system, the spreading effect proceeds
slowly.  When the electron wave packet is scattered by the
double-dot system, the forward scattered wave packet exhibits faster
spreading in the longitudinal direction manifesting a long-tail
behavior caused by the slow release of the probability by the
long-lived resonance state.

\begin{figure*}[htb]
     \includegraphics[width=0.35\linewidth]{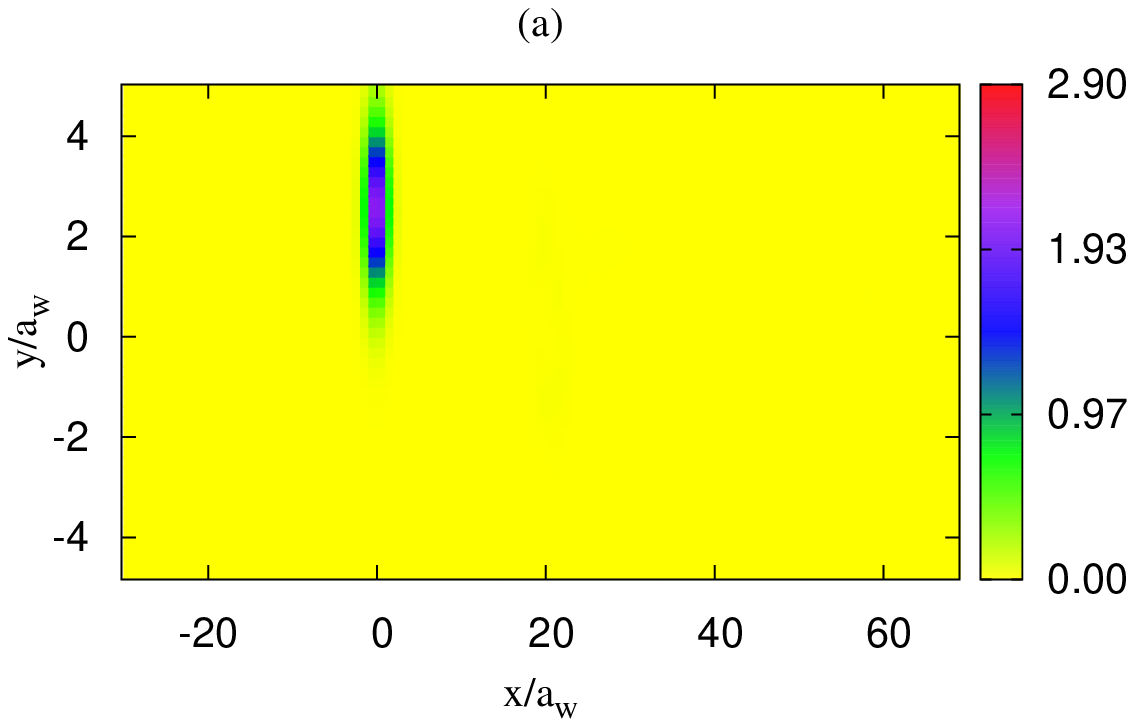}
     \includegraphics[width=0.35\linewidth]{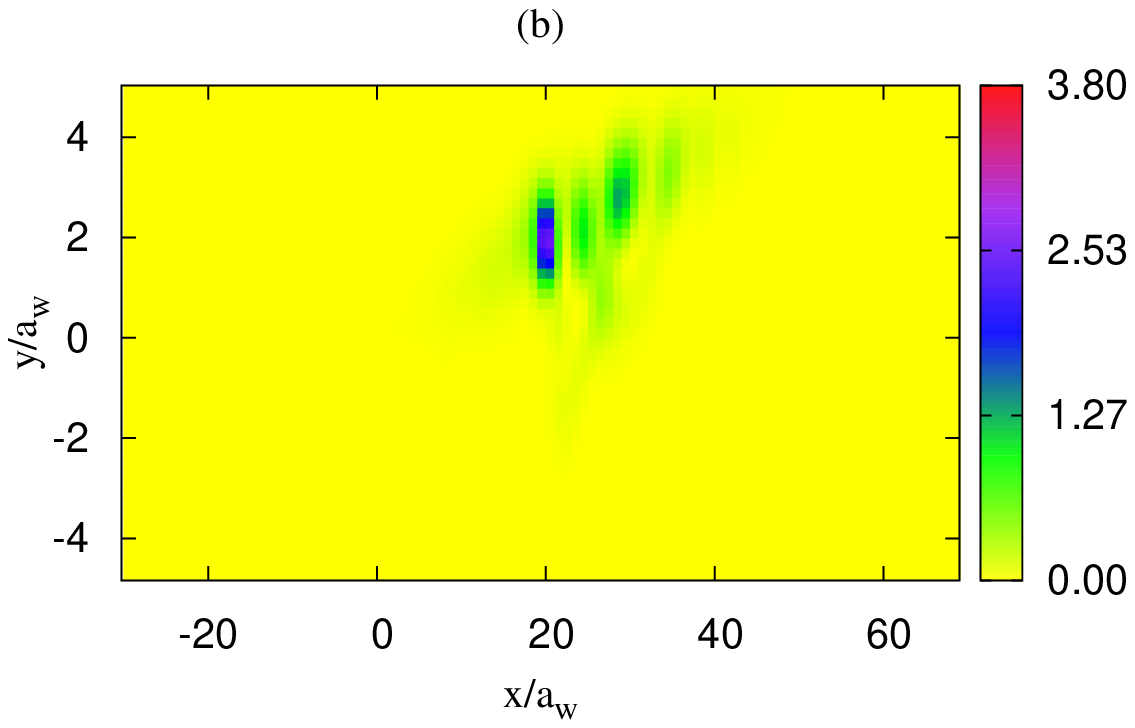}
     \includegraphics[width=0.35\linewidth]{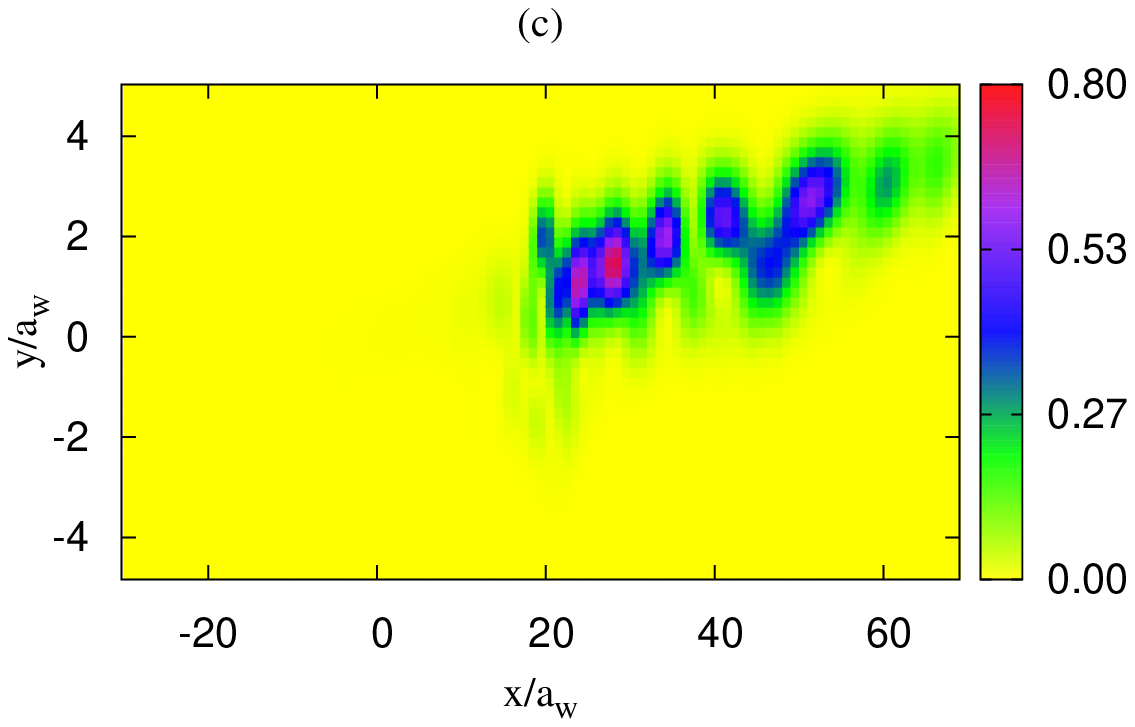}
     \includegraphics[width=0.35\linewidth]{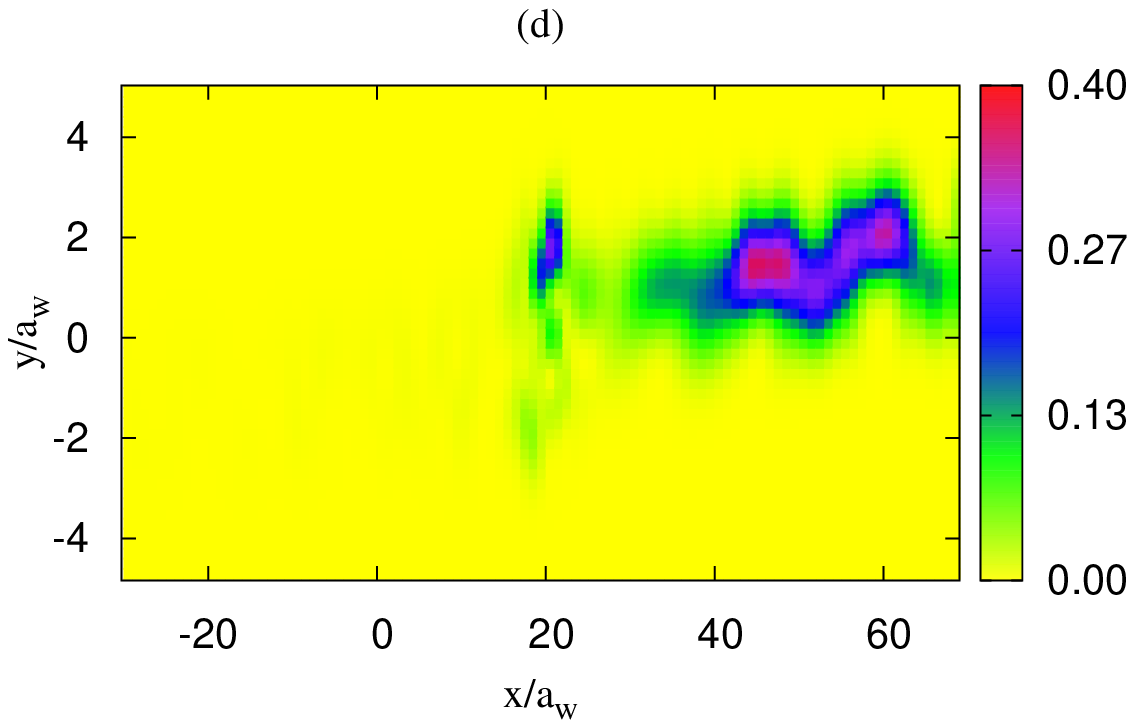}
     \caption{(color online). Propagation of the electron wave packet
     traveling through an embedded parallel double-dot system for
     the case of $B=1.0$~T at the time $t=$ (a) 0; (b) 11; (c) 21; (d) 40~ps.
     The other parameters are the same as \fig{sysDD}.}
     \label{DD-1T}
\end{figure*}

In \fig{DD-05T}(c), at the time 25~ps, part of the wave packet is
clearly coupled to the lower dot although part of the wave packet
has been forward scattered showing skipping-like trajectories and
traveling into the right asymptotic region.  The skipping-like
behavior implies a significant intersubband mixing due to the broad
wave packet distribution in the momentum space.  At $t=38$~ps, the
longitudinal distribution of the forward scattered wave packet is
getting broader; and the localized part remains in the double-dot
region with approximately a half packet height of the localized
state formed at $t=9$~ps as shown in \fig{DD-05T}(b).  We would like
to mention in passing that the reflection wave packet takes more
time to emerge than the transmission wave packet due to the
formation of the quasibound states in the double-dot embedded
system.

We show, in \fig{DD-1T}, that the snapshots of the electron wave
packet propagation through an embedded parallel double-dot system
for the case of $B=1.0$~T at the time $t=$ (a) 0; (b) 11; (c) 21;
(d) 40~ps.  It is shown in \fig{DD-1T}(a) that, at time $t=0$~ps,
the incident wave packet has a compact longitudinal distribution
$\Delta x_{\rm in}=\sqrt{2}a_{\rm w}$ with probability density
height 1.2. A clear localized quasibound state feature is found at
$t=$9~ps as is shown in \fig{DD-1T}(b).  The height of the
probability density of the localized state is twice the height of
the initial wave packet.  At this moment, the backward reflection is
blocked and the forward transmission is relatively low.

The wave packet propagation at the moment $t=21$~ps is depicted in
\fig{DD-1T}(c).  The forward scattered wave packet exhibits rich and
robust transport behavior.  Not only the skipping-like wave packet
flight is found, but also an interference feature attributed to the
intersubband mixing is significantly manifested.  At time $t=40$~ps
shown in \fig{DD-1T}(d), the skipping-like trajectory is still
significant but the interference feature is suppressed.  In
addition, the inter-dot coupling of the wave packet in $B=1.0$~T is
weaker than that in $B=0.5$~T due to the stronger Lorentz force
enhancing off-center shifting.  The localized state for the case of
$B=0.5$~T is covering two quantum dots whereas the localized state
for the case of $B=1.0$~T is mainly in the upper dot.  This implies
that the $0.5$~T magnetic field fits better to the length scales of
the double-dot system.

\section{Concluding remarks}\label{Conclusion}

In this work we have developed a theoretical model by implementing
the Lippmann-Schwinger formalism to demonstrate and elucidate the
transport properties of a Gaussian-type electron wave packet
traveling through a quantum wire with embedded quantum dots under a
homogeneous perpendicular magnetic field.  The magnetic field
induces Lorentz force, which enriches the dynamics of electron wave
packet propagation.  We have found that quantum skipping-like
oscillation trajectory of a wave packet is induced in an appropriate
magnetic field when the wave packet envelop function covers the
lowest two subbands.  This is a quantum forerunner to the well known
skipping orbit motion of classical particles.

For the case of an embedded antidot, the electron wave packet has
been considered with momentum envelop covering the lowest two
subbands.  The wave packet propagation exhibiting non-skipping-like
trajectories implies that the significance of the wave packet
propagation is staying in the lowest subband.  The part of the wave
packet with high kinetic tends to go through the antidot system but
the the part with low kinetic energy is backscattered by the
scattering region.  Quasibound state features with negative binding
energies have also been seen to play an active role in the
scattering process.

For the case of embedded double quantum dot, we have found a robust
trapping effect of the electronic wave packet moving into the
double-dot system and forming localized states.  If there are
several bound states, the electrons make multiple scattering in the
coupled double quantum dot resulting in a superposition of these
bound states, exhibiting oscillating behavior in the double-dot
system.  The parallel double-dot system enables the electron wave
packet performing resonant coupling between two dots in an
appropriate magnetic field and then allows electron wave packet
performing inter-edge backscattering.

The coherent motion of electron waves through open nanostructures in
a penetrating magnetic field may offer promising approaches to
semiconductor spintronics\cite{Zozoulenko04} and controlling the
dynamics of coherent quantum states for quantum information
processing.\cite{Harris01}  To explore these new directions, we need
to track the motion of electron waves in an applied magnetic field.
The cooled scanning probe microscope renders the possibility of
imaging the electron wave trajectories by using the scanning tip as
a movable gate.\cite{Topinka00}  Very recently, quantum dot embedded
mesoscopic system has been utilized for the coherent probing of
excited quantum dot states.\cite{Sigrist07}  We hope that our paper
will stimulate experimental interest to nanostructure embedded
quantum systems in the strong coupling regime, which may provide a
useful tool for the dynamical quantum manipulation of charged
carriers.

\begin{acknowledgments}
      The authors acknowledge financial support from the Research
      and Instruments Funds of the Icelandic State,
      the Research Fund of the University of Iceland, the
      Icelandic Science and Technology Research Programme for
      Postgenomic Biomedicine, Nanoscience and Nanotechnology, and the
      National Science Council of Taiwan.
      C.S.T. is grateful to the computational facilities supported
      by the National Center for High-Performance Computing in Taiwan
      and the University of Iceland.
\end{acknowledgments}


\begin{thebibliography}{99}

% nanotechnology

\bib{nanostructure} See, for example, M. Law, J. Goldberger, and P. Yang,
Annu. Rev. Mater. Res. {\bf 34}, 83 (2004); J. Noborisaka, J.
Motohisa, S. Hara, and T. Fukui, Appl. Phys. Lett. {\bf 87}, 093109
(2005).

% conductance quantization

\bib{Wees88-Koester93-Scappucci06} B. J. van Wees, H. van Houten,
C. W. J. Beenakker, J. G. Williamson, L. P. Kouwenhoven, D. van der
Marel, and C. T. Foxon, Phys. Rev. Lett. {\bf 60}, 848 (1988); S. J.
Koester, C. R. Bolognesi, M. J. Rooks, E. L. Hu, and H. Kroemer,
Appl. Phys. Lett. {\bf 62}, 1373 (1993); G. Scappucci, L. Di
Gaspare, E. Giovine, A. Notargiacomo, R. Leoni, and F. Evangelisti,
Phys. Rev. B {\bf 74}, 035321 (2006).

% static scatterer

\bib{Chu89-Bagwell90-Bardarson04} C. S. Chu and R. S. Sorbello,
Phys. Rev. B {\bf 40}, 5941 (1989); P. F. Bagwell, {\it ibid.}\ {\bf
41}, 10354 (1990);  J. H. Bardarson, I. Magnusdottir, G.
Gudmundsdottir, C. S. Tang, A. Manolescu, and V. Gudmundsson, {\it
ibid.}\ {\bf 70}, 245308 (2004).


% magnetic field

\bibitem{Berggren89-Tang-Vidar} J.~F. Weisz and K.-F. Berggren, Phys. Rev. B
\textbf{40}, 1325 (1989); C. S. Tang and V. Gudmundsson, {\it
ibid.}\ {\bf 74}, 195323 (2006); V. Gudmundsson and C. S. Tang, {\it
ibid.}\ {\bf 74}, 125302 (2006).

\bib{Olendski05} O. Olendski, L. Mikhailovska, Phys. Rev. B {\bf 72}, 235314
(2005).

\bib{Rogge06} M. C. Rogge, F. Cavaliere, M. Sassetti, R. J. Haug, and B. Kramer,
New J. Phys. {\bf 8}, 298 (2006).

% Utilize magnetic field

\bib{Koonen00} J. J. Koonen, H. Buhmann, and L. W. Molenkamp, Phys. Rev.
Lett. {\bf 84}, 2473 (2000).

\bib{Lofgren06} A. L\"{o}fgren, C. A. Marlow, T. E. Humphrey, I.
Shorubalko, R. P. Taylor, P. Omling, R. Newbury, P. E. Lindelof, and
H. Linke, Phys. Rev. B {\bf 73} 235321 (2006).

\bib{Aidala07} K. E. Aidala, R. E. Parrott, T. Kramer, E. J. Heller,
R. M. Westervelt, M. P. Hanson, and A. C. Gossard, Nature Phys. {\bf
3}, 464 (2007).

\bib{Gabelli06-07} J. Gabelli, G. F\`{e}ve, J.-M. Berroir, B. Placais, A. Cavanna, B.
Etienne, Y. Jin, D. C. Glattli, Science {\bf 313}, 499 (2006); J.
Gabelli, G. F\`{e}ve, T. Kontos, J.-M. Berroir, B. Placais, D. C.
Glattli, B. Etienne, Y. Jin, and M. B\"{u}ttiker, Phys. Rev. Lett.
{\bf 98}, 166806 (2007).


%%%%%%   time-dependent transport   %%%%%%
%\bib{Bagwell92} P. F. Bagwell, R. K. Lake, Phys. Rev. B {\bf 46}, 15329 (1992).

\bibitem{TangChu-Wu06} C.~S. Tang and C.~S. Chu, Phys. Rev. B {\bf 53}, 4838 (1996);
{\it ibid.}\ Physica B {\bf 254}, 178 (1998); {\it ibid.}\ Phys.
Rev. B {\bf 60}, 1830 (1999); {\it ibid.}\ Physica B {\bf 292}, 127
(2000); C.~S. Tang, Y.~H. Tan, and C.~S. Chu, Phys. Rev. B {\bf 67},
205324 (2003); B. H. Wu and J. C. Cao, Phys. Rev. B {\bf 73}, 245412
(2006).

% nonadiabatic quantum charge pumping

\bibitem{Tang01-Bylander05} C. S. Tang and C. S. Chu, Solid State Commun. \textbf{120}, 353 (2001);
S. W. Chung, C. S. Tang, C. S. Chu, and C. Y. Chang, Phys. Rev. B
{\bf 70}, 085315 (2004); J. Bylander, T. Duty, and P. Delsing,
Nature {\bf 434}, 361 (2005).

\bib{Blumenthal07} M. D. Blumenthal, B. Kaestner, L. Li, S. Giblin,
T. J. B. M. Janssen, M. Pepper, D. Anderson, G. Jones, and D. A.
Ritchie, Nature Phys. {\bf 3}, 343 (2007).

% current-driven oscillation

\bib{Malshukov05-Kaun05-Pistolesi06} A. G. Mal'shukov, C. S. Tang, C. S. Chu, and K. A. Chao, Phys.
Rev. Lett. {\bf 95}, 107203 (2005); C. C. Kaun and T. Seideman, {\it
ibid.}\ {\bf 94}, 226801 (2005); F. Pistolesi and R. Fazio, New J.
Phys. {\bf 8}, 113 (2006).

\bib{Szafran05} B. Szafran and F. M. Peeters, Phys. Rev. B {\bf 72}, 165301
(2005).

\bib{Okunishi07-Luan07} T. Okunishi, Y. Ohtsuka, M. Muraguchi, and K. Takeda, Phys.
Rev. B {\bf 75}, 245314 (2007); W. H. Kuan, C. S. Tang, and C. H.
Chang, {\it ibid.}\ {\bf 75}, 155326 (2007); P. G. Luan and C. S.
Tang, J. Phys.: Condens. Matter {\bf 19}, 176224 (2007).

%%%%%%   MODEL   %%%%%%

%% momentum-coordinate mixed representation

\bibitem{Gurvitz95-Gudmundsson05} S.~A. Gurvitz, Phys. Rev. B \textbf{51}, 7123
(1995); V. Gudmundsson, Y. Y. Lin, C. S. Tang, V. Moldoveanu, J. H.
Bardarson, and A. Manolescu, Phys. Rev. B {\bf 71}, 235302 (2005).

% Lippmann-Schwinger

\bib{DiVentra01} See, e.g., M. Di Ventra and N. D. Lang, Phys. Rev. B {\bf 65}, 045402
(2001).

% Landauer-Buttiker

\bibitem{Landauer-Buttiker} M. B\"{u}ttiker, Y. Imry, R. Landauer, and S. Pinhas,
Phys. Rev. B {\bf 31}, 6207 (1985); Y. Imry and R. Landauer, Rev.
Mod. Phys. {\bf 71}, S306 (1999).

%%%%%%   Results   %%%%%%

% Antidot

\bib{Sanchez04} D. Sanchez and M. B\"{u}ttiker, Phys. Rev. Lett.
{\bf 93}, 126802 (2004).

\bib{Kirczenow94-Kataoka02} G. Kirczenow, A. S. Sachrajda, Y. Feng,
R. P. Taylor, L. Henning, J. Wang, P. Zawadzki, and P. T. Coleridge,
Phys. Rev. Lett. {\bf 72}, 2069 (1994); M. Kataoka, C. J. B. Ford,
M. Y. Simmons, and D. A. Ritchie, {\it ibid.}\ {\bf 89}, 226803
(2002).

% negative binding energy

\bibitem{Vidar05}  V. Gudmundsson, C.~S. Tang, and A. Manolescu, Phys. Rev. B
{\bf 72}, 153306 (2005).

% quantum difussion

\bib{Zhong01} J. Zhong, R. B. Diener, D. A. Steck, W. H.
Oskay, M. G. Raizen, E. W. Plummer, Z. Zhang, and Q. Niu, Phys. Rev.
Lett. {\bf 86}, 2485 (2001).

%%%%%%   Double Dot   %%%%%%

% DD in series

\bibitem{vdWiel03} For a general overview see, e.g., W.~G. van der Wiel,
S.~D. Franceschi, J.~M. Elzerman, T. Fujisawa, S. Tarucha, L.~P.
Kouwenhoven, Rev. Mod. Phys. {\bf 75}, 1 (2003).

% DD in parallel

\bibitem{Chen04} J.~C. Chen, A.~M. Chang, and M.~R. Melloch, Phys.
Rev. Lett. {\bf 92}, 176801 (2004).

% entanglement

\bib{Loss00-Schomerus07} D. Loss and E. V. Sukhorukov, Phys. Rev. Lett. {\bf 84}, 1035
(2000); H. Schomerus and J. P. Robinson, New J. Phys. {\bf 9}, 67
(2007).

% single electron

\bib{Petta04} J. R. Petta, A. C. Johnson, C. M. Marcus, M. P. Hanson, and
A. C. Gossard, Phys. Rev. Lett. {\bf 93}, 186802 (2004).

% Dephasing

\bibitem{Elhassan05} M. Elhassan, J.~P. Bird, R Akis, D.~K. Ferry,
T. Ida, and K. Ishibashi, J. Phys.: Condens. Matter {\bf 17}, L351
(2005).

% nonadiabatic transport under irradiation

\bib{Naber06} W. J. M. Naber, T. Fujisawa, H. W. Liu, and W. G. van der
Wiel, Phys. Rev. Lett. {\bf 96}, 136807 (2006).

% readout

\bib{Jordan06} A. N. Jordan, A. N. Korotkov, and M. B\"{u}ttiker,
Phys. Rev. Lett. {\bf 97}, 026805 (2006).

% interdot coupling

\bib{Craig04} N. J. Craig, J. M. Taylor, E. A. Lester, C. M. Marcus, M. P.
Hanson, and A. C. Gossard, Science {\bf 304}, 565 (2004).

% Anderson model

\bibitem{Kiselev03-Hartmann04-Zarand06} M. N. Kiselev, K. Kikoin, and L. W. Molenkamp,
Phys. Rev. B {\bf 68}, 155323 (2003); U. Hartmann and F.~K. Wilhelm,
{\it ibid.}\ {\bf 69}, 161309(R) (2004); G. Zar\'{a}nd, C.-H. Chung,
P. Simon, and M. Vojta, Phys. Rev. Lett. {\bf 97}, 166802 (2006).

\bib{Tang05} C. S. Tang, W. W. Yu, and V. Gudmundsson, Phys. Rev. B {\bf 72},
195331 (2005).

%%%%%%   Conclusion   %%%%%%

% spin application

\bib{Zozoulenko04} I. V. Zozoulenko and M. Evaldsson, Appl. Phys.
Lett. {\bf 85}, 3136 (2004).

% qubit

\bib{Harris01} J. Harris, R. Akis, and D. K. Ferry, Appl. Phys.
Lett. {\bf 79} 2214 (2001).

% scanning probe microscope

\bib{Topinka00} M. A. Topinka,  B. J. LeRoy,  S. E. J. Shaw,  E. J. Heller,
R. M. Westervelt,  K. D. Maranowski,  A. C. Gossard, Science {\bf
289}, 2323 (2000); M. A. Topinka, R. M. Westervelt, and E. J.
Heller, Phys. Today {\bf 56}, 47 (2003).

% embedded dot experiment

\bib{Sigrist07} M. Sigrist, T. Ihn, K. Ensslin, M. Reinwald, and
W. Wegscheider, Phys. Rev. Lett. {\bf 98}, 036805 (2007).

\end{thebibliography}
\end{document}